\documentclass[12pt] {iopart}

 \expandafter\let\csname equation*\endcsname\relax
  \expandafter\let\csname endequation*\endcsname\relax
  
\usepackage{amsmath}
\usepackage{iopams}  

\begin{document}

\title[Classical limit of the canonical partition function]{Classical limit of the canonical partition function}

\author{Pere Seglar and Enric P\'{e}rez}

\address{Universitat de Barcelona, Departament de F\'isica Fonamental, c. Mart\'i i Franqu\`es 1, Barcelona 08028}
\ead{enperez@ub.edu}

\begin{abstract}
We analyze the so-called classical limit of the quantum-mechanical canonical partition function. In order to do that, we define accurately the density matrix for symmetrized and antisymmetrized wave functions only (Bose-Einstein and Fermi-Dirac), and find an exact relation between them and the density matrix for non symmetrized functions (Maxwell-Boltzmann). Our results differ from the generally assumed in a numerical factor $N!$, for which we suggest a physical interpretation. We derive as well the reverse (and also exact) relation, to find the canonical partition function for non-symmetrized wave functions in terms of the corresponding function for fermions and bosons. 
\end{abstract}

\maketitle

\section{Introduction}

It is widely known that within the frame of Quantum Mechanics the way of calculating the partition function, cornerstone of Statistical Mechanics, must be drastically changed with respect to Classical Mechanics \cite{Pathria, Huang}. The phase space does not determine the sample space any more, and it is necessary to develop the formalism of the density matrix. A link between both treatments can be traced with the classical limit of the canonical partition function.

	In this paper we are going to perform this limit in detail, focusing our attention on the definition of the few basic elements involved. Our aim is to try to remove some ambiguities in the usual derivations and to offer a new method which suggests a new insight on the subject. As a consequence, we obtain a result, \eqref{starringo}, which differs from the usual one in a numerical factor $N!$, which has been (and still is) the center of much debate over the years \cite{Monaldi}. In the final section we propose a possible physical interpretation for this result. We also provide an exact expression for the inverse relation, between the canonical partition function of symmetrized and non-symmetrized wave functions, \eqref{reves}.

Be that as it may, though we comment on some conceptual issues, what we present here is a mathematical result. 	Our paper only requires elemental calculus of Quantum Mechanics and Statistical Mechanics. Hence, any general physicist should be able to read it, as well as any graduate student. However, it is specially directed to teachers of Statistical Mechanics. We firmly believe that the discussion we propose of the elementary principles of Quantum Statistical Mechanics can be extremely useful both for students and teachers. 

Before starting, we must comment on the notation. As it will be immediately apparent for the reader, our paper contains many indices, subindices and asterisks, and in a couple of occasions we even propose slight provisional changes to make the reading easier. However, despite this apology, we must also remark that, in certain sense, the problem we deal with is a problem of notation. It is the careful treatment of indices and their meaning what allows us to present a new insight on the subject.


\section {Statistics of identical particles. Definitions}

The density matrix $\hat{\rho}_{o}$ is an hermitian operator which can be written, in energy representation, as:

\begin{equation} \label{dmx} 
\hat{\rho}_{o} \equiv \frac{e^{-\beta \hat{H}}}{Tr\{e^{-\beta \hat{H}}  \}}=  
\frac{\displaystyle{\sum_{\underset{\mbox{\scriptsize  \textit{states}}}{\mbox{\scriptsize \textit{all}}}} e^{-\beta E_i} \left. |\psi_{E_i} \right\rangle \left\langle \psi_{E_i}| \right.}}
{Tr \left\{ \displaystyle{ \sum_{\underset{\mbox{\scriptsize  \textit{states}}}{\mbox{\scriptsize \textit{all}}}}} e^{-\beta E_i}  \left. |\psi_{E_i} \right\rangle \left\langle \psi_{E_i}|  \right. \right\}} ,
\end{equation}
where $\beta=1 / \kappa T$ and $\kappa$ is Boltzmann's constant. $\hat{H}$ represents the hamiltonian operator and the sum is extended along all the eigenfunctions $\psi_{E_i}$ of energy. These $\psi_{E_i}$ are solutions of the Schr\"odinger equation; using coordinate representation (in order to avoid a clumsy notation we will write $(1, \cdots, N)$ instead of $(\vec{q}_1, \cdots, \vec{q}_N)$):

\begin{equation}  \label{ES}
\hat{H} (1, \cdots, N)\psi_{E_i}(1, \cdots,N)=E_i \psi_{E_i}(1, \cdots,N).
\end{equation}
The subindex $E_{i}$ must be understood as representing the set of quantum numbers corresponding to this state. Of course, the level $E_i$ can be degenerated. For the equality in \eqref{dmx} to be true, the $ \left. |\psi_{E_i}\right\rangle $'s have to be a complete set of normalized solutions, and therefore fulfill the identity relation: 

\begin{equation}  \label{identical}
 \hat{I}= \sum_{\underset{\mbox{\scriptsize  \textit{states}}}{\mbox{\scriptsize \textit{all}}}} \left. |\psi_{E_i} \right\rangle \left\langle \psi_{E_i}| \right. .
 \end{equation}

We are only interested in the quantum-mechanical partition function, which is defined as the normalization factor in the denominator of \eqref{dmx}:

$$Z(\beta, V,N) \equiv Tr \{e^{-\beta \hat{H}}  \}. $$
Then, we are going to operate with the matrix elements of $\hat{\rho}_o$ without the normalization, which we will call $\hat{\rho}$. That is:

$$\hat{\rho} \equiv e^{-\beta \hat{H}},$$

$$\hat{\rho}_{o} \equiv \frac{\hat{\rho}}{Tr\{\hat{\rho} \}},$$
and

$$Tr\{ \hat{\rho}_o  \}=1.$$

The matrix elements of $\hat{\rho}$ are:

\begin{equation} \label{dmxe}
\begin{array} {c}
\langle 1,\cdots,N | \hat{\rho} | 1', \cdots ,N' \rangle \equiv \hat{\rho}(1,\cdots,N;1',\cdots,N')= \\ =\displaystyle{\sum_{\underset{\mbox{\scriptsize  \textit{states}}}{\mbox{\scriptsize \textit{all}}}}} 
 \psi^{*}_{E_i}(1',\cdots,N') e^{-\beta \hat{H}(1, \cdots, N)} \psi_{E_i}(1,\cdots,N) =\\=\displaystyle{\sum_{\underset{\mbox{\scriptsize  \textit{states}}}{\mbox{\scriptsize \textit{all}}}}} 
\psi^{*}_{E_i}
(1',\cdots, N')e^{-\beta E_i} \psi_{E_i}(1,\cdots,N)=\\=\displaystyle{\sum_{\underset{\mbox{\scriptsize  \textit{states}}}{\mbox{\scriptsize \textit{all}}}}} 
e^{-\beta E_i} \psi_{E_i}(1,\cdots,N)\psi^{*}_{E_i}
(1',\cdots, N').
\end{array}
\end{equation}
%

The solutions $\psi_{E_i}$ do not necessarily possess any symmetry property. In fact, in general, they do not, even if the hamiltonian is symmetric. For a certain system, some of the solutions are symmetric (or antisymmetric) and some of them are not. However, due to the indistinguishability of particles we must allow only symmetric (or antisymmetric) functions. 

To symmetrize (\textit{S}) or antisymmetrize (\textit{A}) functions starting from solutions in \eqref{ES} we will follow the usual method of constructing symmetrized (antisymmetrized) functions  through linear combinations of members of the basis with the same energy $E_i$: 

\begin{eqnarray} \label{psisa}
\psi^{S/A}_{E_i}(1, \cdots, N) \equiv \frac{1}{\sqrt{N!}} \sum_{P} (\pm 1)^{P} \psi_{E_i} (P[1, \cdots, N])
\end{eqnarray} 
($P$ indicates permutations among coordinates). Note that \eqref{psisa} is a definition, and that all the functions in the right hand are eigenfunctions of the same value of energy $E_i$, but are still different (this is the so-called interchange degeneration). In fact, they are orthogonal. Of course, this is not the case if they are originally symmetric/antisymmetric.

\subsection{Hilbert subspaces}

Let us exemplify the constructing rule in \eqref{psisa} with the 2-particle system. In this case we have:

 \begin{equation} \label{syminv}
\psi^{S/A}_{E_i} (1, 2) \equiv \frac{1}{\sqrt{2}} \left[\psi_{E_i} (1, 2) \pm \psi_{E_i}(2, 1) \right],
\end{equation}
and we can easily invert the relation writing an identity equality:

\begin{equation} \label{desc}
\psi_{E_i}(1,2)=\displaystyle{\frac{1}{2}} \left\{ \left[\psi_{E_i}(1, 2)+\psi_{E_i}(2,1)\right] + \left[\psi_{E_i}(1,2)-\psi_{E_i}(2, 1)\right] \right\},
\end{equation}
and then noting that: 

\begin{equation} \label{desc}
\psi_{E_i}(1,2)=\displaystyle{\frac{1}{\sqrt{2}}} \left[\psi_{i}^{S}(1,2)+ \psi_{i}^{A}(1,2)\right].
\end{equation}

The 2-particle system constitutes a very special case, which makes it qualitatively different from any other system with $N>2$. In this sense, it is not a good example. We have seen that for $N=2$ we can write any $\psi_{E_i}(1,2)$ in terms of symmetrized and antisymmetrized functions only. According to group theory, this is a consequence of the fact that the Hilbert space of two particles $\mathcal{H}^{(2)}$ decomposes in 2 subspaces, the symmetric and the antisymmetric:

$$\mathcal{H}^{(2)}=\mathcal{H}_S \oplus \mathcal{H}_A.$$
Each subspace corresponds to one of the two eigenvalues of operator permutation $\hat{P}$ ($+1$ and $-1$), which satisfies $[\hat{H},\hat{P}]=0$. Therefore, there are two eigenfunctions of the same energy $E_i$ which differ only in one permutation.

For the general case, it is known that the Hilbert space must be decomposed into many subspaces, not only two \cite{Galindo, Hamermesh}:

\begin{equation} \label{spatiald}
\mathcal{H}^{(N)}=\mathcal{H}_{S}\oplus \mathcal{H}_{A} \oplus \sum_{\underset {s_j}{\underset{\mbox{\scriptsize  \textit{symmetries}}}{\mbox{\scriptsize \textit{mixed}}}}} \mathcal{H}_{s_j}.
\end{equation}
These Hilbert subspaces are generated by symmetrized functions, antisymmetrized functions, and functions with other symmetries. The latter are not uniquely defined, which means that we can choose different generators of these subspaces. Anyway, any operation of symmetrization characteristic of each group (symmetrize, antisymmetrize, or the symmetrization rules corresponding to the other subspaces) applied to a eigenfunction not belonging to its own subspace yields zero. They are completely disjoint sets. 

In terms of the wave function the decomposition means:

\begin{equation} 
\psi_{E_i} (1,\cdots,N)=a_{S}\psi_{E_i}^{S}(1,\cdots,N)+b_{A}\psi_{E_i}^{A}(1,\cdots,N)+ \sum_{\underset {s_j}{\underset{\mbox{\scriptsize  \textit{symmetries}}}{\mbox{\scriptsize \textit{mixed}}}}} c_{s_j} \psi_{E_i}^{s_{j}}(1,\cdots,N).
\label{dcpm}
\end{equation}
We can easily derive the coeficients $a_{S}$ and $b_{A}$ performing a permutation. According to \eqref{psisa}:

$$\sum_{P} (\pm1)^{P}\psi_{E_i} (P[1,\cdots,N])=\sqrt{N!} \psi_{E_i}^{S/A}(1,\cdots,N),$$
and due to the aforementioned properties of Hilbert subspaces, we have, for the function \eqref{dcpm}:

$$\sum_{P} \psi_{E_i} (P[1,\cdots,N])=a_{S}N! \psi_{E_i}^{S}(1,\cdots,N).$$
And also:

$$\sum_{P} (-1)^{P} \psi_{E_i} (P[1,\cdots,N])=b_{A}N! \psi_{E_i}^{A}(1,\cdots,N).$$
Then:

\begin{equation}  \label{coeff}
a_{S}=b_{A}=\frac{1}{\sqrt{N!}}.
\end{equation}
And finally we can write:
\begin{equation} \label{dcpmm}
\begin{array} {c}
\psi_{E_i} (1,\cdots,N)=\displaystyle{\frac{1}{\sqrt{N!}} }\left\{ \psi_{E_i}^{S}(1,\cdots,N)+ \psi_{E_i}^{A}(1,\cdots,N) \right\} + \sum_{\underset {s_j}{\underset{\mbox{\scriptsize  \textit{symmetries}}}{\mbox{\scriptsize \textit{mixed}}}}} c_{s_j} \psi_{E_i}^{s_{j}}(1,\cdots,N).
\end{array}
\end{equation}
Note what this expression says: the set built with symmetrized and antisymmetrized functions is not complete. This is logical, as long as there are functions with no symmetry properties in the original (and complete) basis.

In the Appendix A we give the coefficients $c_{s_j}$ for the 3-particle system.


\section{Canonical partition function of symmetrized functions}

\subsection{The density matrix}

Let us now go back to calculate the density matrix (without normalization) for systems of identical particles. Due to the requirements of indistinguishability, we must work with the symmetrized functions in \eqref{psisa}. First of all, we define, in analogy with \eqref{dmxe}, the matrix elements in coordinate representation:\footnote{Note that due to the fact that the symmetrized functions do not constitue a complete set of solutions, we cannot write an analogous expression to \eqref{dmx}.}

\begin{equation} \label{dmx2}
\hat{\rho}_{S/A} (1,\cdots,N;1',\cdots,N')\equiv \sum_{\underset{\mbox{\scriptsize  \textit{states only}}}{\mbox{\scriptsize \textit{sym./antisym.}}}} e^{-\beta E_i} \psi_{E_i}^{S/A}(1,\cdots,N) \psi_{E_i}^{*S/A}(1',\dots,N') .
\end{equation}

What we want to do in the following is to write these matrix elements in terms of the classical, unsymmetrized ones in \eqref{dmxe}. Let us begin with the 2-particle system. We are going to follow a method very similar to that proposed by Feynman \cite{Feynman}. It consists of noting that, as:

\begin{equation}
\hat{\rho} (1,2;1',2')=\sum_{\underset{\mbox{\scriptsize  \textit{states}}}{\mbox{\scriptsize \textit{all}}}} e^{-\beta E_i} \psi_{E_i}(1,2) \psi^{*}_{E_i}(1',2'),
\end{equation}
then the combination:\footnote{Feynman only permutes the non-prime coordinates.}

\begin{equation} \label{correctav}
\frac{1}{4} \left[ \hat{\rho}(1, 2;1', 2') \pm \hat{\rho}(1, 2;2', 1') \pm \hat{\rho}(2,1;1',2') + \hat{\rho}(2,1;2',1') \right]
\end{equation}
yields:

\begin{equation} \label{correctavvv}
\displaystyle{\sum_{\underset{\mbox{\scriptsize  \textit{states}}}{\mbox{\scriptsize \textit{all}}}}} e^{-\beta E_i} \frac{1}{2} \left[\psi_{E_i}(1,2) \pm \psi_{E_i}(2, 1) \right]  \frac{1}{2} \left[\psi^{*}_{E_i}(1', 2') \pm \psi^{*}_{E_i}(2', 1') \right],
\end{equation}
where we still must sum over \textit{all states}. But \eqref{correctavvv} and \eqref{syminv} yields:

\begin{equation} \label{correctavv}
\displaystyle{\frac{1}{2}}\displaystyle{\sum_{\underset{\mbox{\scriptsize  \textit{states only}}}{\mbox{\scriptsize \textit{sym./antisym.}}}}}  e^{-\beta E_{i} } \psi_{E_i}^{S/A}(1, 2) \psi^{*S/A}_{E_i}(1', 2').
\end{equation}
Note that now, the summation is extended to symmetrized/antisymmetrized functions only. The fact is that, although \eqref{correctavvv} and \eqref{correctavv} are completely equivalent, in general we only know how to perform the sum in the first case. We will come back to this point in Appendix C, for the special case of ideal gases.

Finally, for the 2-particle system, according to \eqref{correctav} and \eqref{correctavv} we can certainly write:

\begin{equation} \label{vinti}
\begin{array}{c}
\hat{\rho}_{S/A}(1, 2; 1', 2')= \displaystyle{\frac{1}{2}} \displaystyle{\sum_{P}}\displaystyle{\sum_{P'}} (\pm 1)^{P+P'} \hat{\rho}(P[1,2]; P'[1',2']) .
\end{array}
\end{equation}

One can reproduce this reasoning for $N$ particles. The generalization of \eqref{vinti} is:

\begin{equation} \label{silkrule}
\hat{\rho}_{S/A}(1, \cdots, N; 1', \cdots, N') = \frac{1}{N!} \sum_{P} \displaystyle{\sum_{P'}}(\pm 1)^{P+P'} \hat{\rho}(P[1, \cdots, N]; P'[1', \cdots, N']).
\end{equation}
This is a general result, not restricted only to ideal gases.

\subsection{The partition function}

In order to obtain the partition function, we have to caculate the trace of $\hat{\rho}_{S/A}$ in \eqref{silkrule}. After suppressing the primes we easily see that we have $N!$ equivalent terms corresponding to the identical permutations. Therefore, the diagonal elements are:

\begin{equation} \label{permi}
\begin{array} {c}
\hat{\rho}_{S/A}(1, \cdots, N;1, \cdots, N) = \displaystyle{\frac{1}{N!}} \sum_{P} \displaystyle{\sum_{P'}}(\pm 1)^{P+P'} \hat{\rho}(P[1, \cdots, N]; P'[1, \cdots, N])=\\=\displaystyle{\sum_{P}} (\pm 1)^{P} \hat{\rho}(1, \cdots, N; P[1, \cdots, N]) .
\end{array}
\end{equation}
And the trace:

\begin{equation} \label{permi2}
\begin{array} {c}
Tr \{\hat{\rho}_{S/A}(1, \cdots, N;1, \cdots, N)\} =\\= \displaystyle{\int \cdots \int}  \sum_{P} (\pm 1)^{P} \hat{\rho}(1, \cdots, N; P[1, \cdots, N]) d1 \cdots dN.
\end{array}
\end{equation}

This is the partition function $Z_{S/A}(\beta, V,N)$ of a system of $N$ identical particles calculated in terms of the density matrix of non-symmetrized wave functions.

We can easily check now the good agreement between \eqref{permi}, \eqref{dmx2} and \eqref{dcpmm}. If we substitute \eqref{dcpmm} into the right side of \eqref{permi}, due to the symmetry properties we immediately obtain that:

\begin{equation} \label{resopo}
\begin{array} {c}

\displaystyle{\sum_{P}} (\pm 1)^{P}\hat{\rho}(1, \cdots,N;P[1, \cdots,N])= \\ = \displaystyle{\sum_{\underset{\mbox{\scriptsize  \textit{states only}}}{\mbox{\scriptsize \textit{sym./antisym.}}}}} e^{-\beta E_i}  \left\{ \psi_{E_i}^{S/A}(1,\cdots,N)\psi_{E_i}^{*S/A}(1,\cdots,N) \right\} +
 \\ + \displaystyle{\sum_{\underset{\mbox{\scriptsize  \textit{states only}}}{\mbox{\scriptsize \textit{sym. \& antisym.}}}}} e^{-\beta E_i} \left\{ \psi_{E_i}^{A/S}(1,\cdots,N)\psi_{E_i}^{*S/A}(1,\cdots,N) \right\} + \\+    \displaystyle{\sum_{\underset{\mbox{\scriptsize  \textit{states only}}}{\mbox{\scriptsize \textit{sym./antisym. \& mix. sym.}}}}}  e^{-\beta E_i} \left\{ \left[\displaystyle{\sum_{\underset{s_j}{\underset{\mbox{\scriptsize  \textit{symm.}}}{\mbox{\scriptsize \textit{mixed}}}}}}  c_{s_j} \psi_{E_i}^{s_j}(1,\cdots,N) \right] \displaystyle {\frac{\psi_{E_i}^{*S/A}(1,\cdots,N)}{\sqrt{N!}}} \right\}.

\end{array}
\end{equation}
Note that the summations have to be extended along diferent domains diferent depending on which subspace of \eqref{spatiald} is involved. If we calculate the trace we get:

$$ \int \cdots \int \displaystyle{\sum_{\underset{\mbox{\scriptsize  \textit{states only}}}{\mbox{\scriptsize \textit{sym./antisym.}}}}} e^{-\beta E_i} \left\{\psi_{E_i}^{S/A}(1,\cdots,N)\psi_{E_i}^{*S/A}(1,\cdots,N)\right\}d1 \cdots dN ,$$
which is exactly the definition of $Tr\{ \hat{\rho}_{S/A} \}$ in \eqref{dmx2}.

\subsection{The classical limit}

Finally, to obtain the classical limit we have to deal with the partition function in \eqref{permi2}:

\begin{equation} \label{permi3}
Z_{S/A}(\beta, V,N) =   \sum_{P} (\pm 1)^{P} \int \cdots \int \hat{\rho}(1, \cdots, N; P[1, \cdots, N]) d1 \cdots dN. 
\end{equation}
To perform it we have to operate on this expression. First of all, we will write it in terms of the eigenfunctions as in \eqref{dmxe}:

\begin{equation} \label{permi4}
\sum_{P} (\pm 1)^{P} \displaystyle{\sum_{\underset{\mbox{\scriptsize  \textit{states}}}{\mbox{\scriptsize \textit{all}}}}} \int \cdots \int  \psi^{*}_{E_i}(P[1, \cdots, N])  e^{-\beta \hat{H}(1,\cdots,N)} \psi_{E_i}(1, \cdots, N) d1 \cdots dN
\end{equation}
($\hat{H}(1,\cdots,N)$ is in coordinate representation). Now, we will expand $\psi_{E_i}(1,\cdots,N)$ in terms of the eigenfunctions of momenta $\chi(\vec{p_1}, \cdots, \vec{p}_N; \vec{q_1}, \cdots, \vec{q}_N)$, which are also eigenfunctions of the kinetic part of $\hat{H}$. In order to do that, we will change slightly our notation. To begin with, we have to retrieve the $(\vec{q_1}, \cdots, \vec{q}_N)$ instead of $(1, \cdots, N)$. In fact, we will refer to $(q_{1x},q_{1y}, \cdots, q_{Nz})$ as $\vec{\textbf{Q}}$, and $(p_{1x},p_{1y}, \cdots, p_{Nz})$ as $\vec{\textbf{P}}$. According to this:

\begin{equation} \label{traf}
\psi_{E_i}(\vec{\textbf{Q}})=\int \cdots \int \varphi_{E_i} (\vec{\textbf{P}}) \chi(\vec{\textbf{P}};\vec{\textbf{Q}}) d\vec{\textbf{P}},
\end{equation}
where

\begin{equation}   \label{funcprop}
\chi(\vec{\textbf{P}};\vec{\textbf{Q}})= {\left( \frac{2 \pi}{h } \right)}^{\frac{3N}{2}} e^{{\frac{2 \pi i}{h}} \vec{\textbf{P}}\cdot\vec{\textbf{Q}}}.
\end{equation}
The inversion of \eqref{traf} is:

\begin{equation} \label{trafu}
\varphi_{E_i}(\vec{\textbf{P}})= \frac{1}{(2 \pi)^{3N}} \int \cdots \int \psi_{E_i} (\vec{\textbf{Q}}) \chi^{*}(\vec{\textbf{P}};\vec{\textbf{Q}}) d\vec{\textbf{Q}}.
\end{equation}

Substituting \eqref{traf} into \eqref{permi4} yields, for $Z_{S/A}(\beta, V,N) $:

\begin{equation} \label{permi5}
\displaystyle{\sum_{P} (\pm 1)^{P}} \displaystyle{\sum_{\underset{\mbox{\scriptsize  \textit{states}}}{\mbox{\scriptsize \textit{all}}}}} \int \cdots \int \psi_{E_i}^{*}(P[\vec{\textbf{Q}}])    \varphi_{E_i}(\vec{\textbf{P}}) e^{-\beta \hat{H}(1,\cdots,N)} \chi(\vec{\textbf{P}}; \vec{\textbf{Q}}) d\vec{\textbf{P}}d\vec{\textbf{Q}}.
\end{equation}
And then, introducing the inversion \eqref{trafu}:

\begin{equation} \label{trapi}
\frac{1}{(2 \pi)^{3N}}\displaystyle{ \sum_{P} (\pm 1)^{P}} \displaystyle{\sum_{\underset{\mbox{\scriptsize  \textit{states}}}{\mbox{\scriptsize \textit{all}}}}} \int \cdots \int \psi_{E_i}^{*}(P[\vec{\textbf{Q}}]) \psi_{E_i}(\vec{\textbf{Q}}') \chi^{*}(\vec{\textbf{P}};\vec{\textbf{Q}}') e^{-\beta \hat{H}(1,\cdots,N)} \chi(\vec{\textbf{P}};\vec{\textbf{Q}})  d\vec{\textbf{P}} d\vec{\textbf{Q}} d\vec{\textbf{Q}}'.
\end{equation}

Now we can use closure condition (see Appendix B) and sum for all $\psi_{E_i}$'s, and obtain:

\begin{equation}
\displaystyle{\sum_{\underset{\mbox{\scriptsize  \textit{states}}}{\mbox{\scriptsize \textit{all}}}}}  \psi^{*}_{E_i} (P[\vec{\textbf{Q}}]) \psi_{E_i} (\vec{\textbf{Q}}')  = \delta(P[\vec{q_1}] - \vec{q}'_{1}) \cdots \delta(P[\vec{q_{N}}]-\vec{q'_{N}}).
\end{equation}
Therefore, \eqref{trapi} becomes:

\begin{equation}
\begin{array} {c}
\frac{1}{(2 \pi)^{3N}} \displaystyle{ \sum_{P} (\pm 1)^{P}} \int \cdots \int  \chi^{*}(\vec{\textbf{P}}, P[\vec{\textbf{Q}}]) e^{-\hat{H}(1,\cdots,N)} \chi(\vec{\textbf{P}};\vec{\textbf{Q}}) d\vec{\textbf{P}}d\vec{\textbf{Q}} =\\
=\frac{1}{h^{3N}} \displaystyle{ \sum_{P} (\pm 1)^{P}} \int \cdots \int  e^{-\beta \left(\frac{\vec{\textbf{P}}^{2}}{2m}+V(\vec{\textbf{Q}}) \right)} e^{{\frac{2 \pi i}{h}} \vec{\textbf{P}} \cdot (\vec{\textbf{Q}}-P[\vec{\textbf{Q}}])} d\vec{\textbf{P}} d\vec{\textbf{Q}}.
\end{array}
\end{equation}

Coming back to the explicit notation, we can finally write, for $Z_{S/A}(\beta, V,N)$:

$$\displaystyle{\frac{1}{(h)^{3N}} \sum_{P} (\pm 1)^{P}} \int \cdots \int  e^{-\beta \displaystyle{\sum_{l=1}^{N} }\left(\frac{\vec{p_{l}}^{2}}{2m}+V(\vec{q_1}, \cdots, \vec{q_N}) \right)} e^{{\frac{2 \pi i}{h}} \displaystyle{\sum_{l=1}^{N}} \vec{p_{l}} (\vec{q_{l}}-P[\vec{q_{l}}])} d\vec{p}_{1}\cdots d\vec{p}_{N}d\vec{q}_{1} \cdots d\vec{q}_{N}.$$

This is a significant result. We have expressed $Z_{S/A}(\beta, V,N)$ as a sum of integrals in the $\Gamma$-space, and the symmetry, finally, only affects the eigenfunctions of momenta. Up to here, the only approximation we have made is the transition to continuum. This is what we have assumed when adopting as eigenfunctions of momenta the functions in \eqref{funcprop}.

Before performing the classical limit, we can integrate momenta. Recalling the definition of the so-called thermal wavelength

$$\Lambda \equiv \frac{h}{\sqrt{2 \pi m k T}},$$
we can write:

\begin{equation} \label{liu}
Z_{S/A}(\beta, V,N)=\frac{1}{\Lambda^{3N}} \sum_{P} (\pm 1)^{P} \int \cdots \int e^{-\beta V(\vec{q_1}, \cdots, \vec{q_N}) }  e^{-\frac{\pi}{\Lambda^{2}} ({\vec{\zeta}_1}^{2} + \cdots + {\vec{\zeta}_N}^{2})} d\vec{q}_{1} \cdots d\vec{q}_{N},
\end{equation}
where:
$$\vec{\zeta}_k \equiv \vec{q}_k - P[\vec{q}_k].$$

To obtain the classical limit, we have to perform the limit $h \rightarrow 0$, that is, $\Lambda \rightarrow 0$. As a consequence, we can approximate the symmetric/antisymmetric partition function with the first term in \eqref{liu}, because then $\vec{\zeta}_k=0$, and the exponential is unity. That is, the term in which the permutation is identity. The rest of terms are corrections of order of $\Lambda^{2}$\cite{Grossmann}. Finally:

\begin{equation} \label{starring}
Z_{S/A}(\beta, V,N) \approx \frac{1}{\Lambda^{3N}} \int \cdots \int  e^{-\beta V(\vec{q}_1, \cdots, \vec{q}_N)} d\vec{q}_{1} \cdots d\vec{q}_{N}.
\end{equation}
That is:
\begin{equation} \label{starringo}
Z_{S/A}(\beta, V,N) = Z(\beta, V,N) (1 + O(\Lambda^{2})).
\end{equation}
Therefore, both partition function of symmetrized or antisymmetrized functions tend to the classical one. Note that for free particles the order of the correction in lambda will be greater.


\subsection {The reverse relation}

For the sake of completeness, let us now deduce the reverse relation between $\hat{\rho}_{S/A}$ and $\hat{\rho}$ in \eqref{permi}. That is, $Z(\beta, V,N)$ in terms of $Z_{S}(\beta, V,N)$ and $Z_{A}(\beta, V,N)$. From what we have said up to now, it is easy to see that:

\begin{equation}
\hat{\rho} \neq \hat{\rho}_{S} + \hat{\rho}_A.
\end{equation}

This is true even in the special case of the $2$-particle system, where any function is the sum of a symmetric and an antisymmetric functions. Recalling that the matrix elements are:\footnote{We retrieve our original notation: $(1, \cdots, N)$ instead of $(\vec{q}_{1}, \cdots, \vec{q}_N)$.} 

\begin{equation} \label{mtzdsc}
\hat{\rho}(1,2; 1', 2')=\sum_{\underset{\mbox{\scriptsize  \textit{states}}}{\mbox{\scriptsize \textit{all}}}} e^{-\beta E_i} \psi_{E_i}(1,2) \psi^{*}_{E_i}(1',2'),
\end{equation}
and substituting here the expression for $\psi_{E_i}^{S/A}(1,2)$ in \eqref{syminv} we readily obtain:

\begin{equation} \label{dosca}
\hat{\rho} (1, 2;1',2')=\frac{\hat{\rho}_{S}(1, 2;1',2') + \hat{\rho}_{A}(1,2;1',2')}{2} + \hat{\rho}_{M}(1, 2;1'2'),
\end{equation}
where:

$$
\hat{\rho}_S(1,2;1'2')=\displaystyle{\sum_{\underset{\mbox{\scriptsize  \textit{states only}}}{\mbox{\scriptsize \textit{symmetric}}}}} e^{-\beta E_{i}} \psi_{E_i}^{S} (1,2) \psi_{E_i}^{*S} (1',2'),
$$

$$
\hat{\rho}_A(1,2;1'2')=\displaystyle{\sum_{\underset{\mbox{\scriptsize  \textit{states only}}}{\mbox{\scriptsize \textit{antisymmetric}}}}} e^{-\beta E_{i}} \psi_{E_i}^{A} (1,2) \psi_{E_i}^{*A} (1',2'),
$$
and $\hat{\rho_{M}}$ stands for mixed terms:

\begin{equation}
\hat{\rho}_{M} (1, 2;1',2')=  
\displaystyle{\sum_{\underset{\mbox{\scriptsize  \textit{states only}}}{\mbox{\scriptsize \textit{sym. \& antisym.}}}}} e^{-\beta E_i} \left[ \psi_{E_i}^{S}(1,2) \psi^{*A}_{i}(1', 2') + \psi^{A}_{i}(1,2)  \psi^{*S}_{i}(1', 2') \right].
\end{equation}

For the general case, we have to appeal again to the decomposition in \eqref{dcpmm}. Substituting it into \eqref{dmxe} we obtain:

\begin{equation} \label{carro}
\begin{array} {c}

\hat{\rho}(1, \cdots, N; 1', \cdots, N')=\\ =
\displaystyle{\sum_{\underset{\mbox{\scriptsize  \textit{states only}}}{\mbox{\scriptsize \textit{sym. \& antisym.}}}}}  e^{-\beta E_{i}}\frac{1}{N!} \left\{ \psi_{E_i}^{S}(1,\cdots,N)\psi_{E_i}^{*S}(1',\cdots,N') + \psi_{E_i}^{A}(1,\cdots,N)\psi_{E_i}^{*A}(1',\cdots,N')  \right\} + \\ + \displaystyle{\sum_{\underset{\mbox{\scriptsize  \textit{states only}}}{\mbox{\scriptsize \textit{sym. \& antisym.}}}}} e^{-\beta E_{i}} \frac{1}{N!} \left\{ \psi_{E_i}^{S}(1,\cdots,N)\psi_{E_i}^{*A}(1',\cdots,N') +\psi_{E_i}^{A}(1,\cdots,N)\psi_{E_i}^{*S}(1',\cdots,N') \right\} +\\

+ \displaystyle{\sum_{\underset{\mbox{\scriptsize  \textit{states}}}{\mbox{\scriptsize \textit{sym. \& antisym. \& mix. sym.}}}}} e^{-\beta E_{i}} \frac{1}{\sqrt{N!}} \left\{ \left[\psi_{E_i}^{S}(1,\cdots,N)+\psi_{E_i}^{A}(1,\cdots,N) \right] \left[\displaystyle{\sum_{\underset{s_j}{\underset{\mbox{\scriptsize  \textit{symm.}}}{\mbox{\scriptsize \textit{mixed}}}}}}  c_{s_j} \psi_{E_i}^{*s_{j}} (1',\cdots,N') \right]+ \right. \\ \left. +\left[\displaystyle{\sum_{\underset{s_j}{\underset{\mbox{\scriptsize  \textit{symm.}}}{\mbox{\scriptsize \textit{mixed}}}}}} c_{s_j} \psi_{E_i}^{s_{j}}(1,\cdots,N)\right] \left[\psi_{E_i}^{*S}(1',\cdots,N')+\psi_{E_i}^{*A}(1',\cdots,N') \right]  \right\} +\\
+ \displaystyle{\sum_{\underset{\mbox{\scriptsize  \textit{states only}}}{\mbox{\scriptsize \textit{mix. sym.}}}}} e^{-\beta E_{i}} \left[\displaystyle{\sum_{\underset{s_j}{\underset{\mbox{\scriptsize  \textit{symm.}}}{\mbox{\scriptsize \textit{mixed}}}}}} c_{s_j} \psi_{E_i}^{s_{j}}(1,\cdots,N)\right]\left[\displaystyle{\sum_{\underset{s_k}{\underset{\mbox{\scriptsize  \textit{symm.}}}{\mbox{\scriptsize \textit{mixed}}}}}} c_{s_k} \psi_{E_i}^{*s_{k}}(1',\cdots,N')\right].
  \end{array}
\end{equation}

That is: 

\begin{equation} \label{goldenrule}
\hat{\rho}=\frac{\hat{\rho}_{S}+ \hat{\rho}_{A}}{N!}
+ \hat{\rho}_{M}
\end{equation}
($\hat{\rho}_{M}$ stands for mixed terms and symmetries). 

Finally, in order to obtain the canonical partition function we must calculate the trace. To do it, we take the primes out of \eqref{carro} and integrate for all the coordinates. Recalling once more symmetry properties we obtain:

\begin{equation} \label{reves}
Z(\beta, V,N)= \frac{1}{N!} [Z_{S}(\beta, V,N)+Z_{A}(\beta, V,N)]+Z_{M}(\beta, V, N),
\end{equation}
where $Z_{M}$ stands for the part corresponding to mixed symmetries: 

\begin{equation}  \label{purria}
\begin{array} {c}
Z_{M}(\beta, V,N)= Tr\left\{\hat{\rho} _{M}(1, \cdots , N;1, \cdots, N)  \right\}= \\ = \displaystyle{\sum_{\underset{\mbox{\scriptsize  \textit{states only}}}{\mbox{\scriptsize \textit{mix. sym.}}}}} \displaystyle{ \int \cdots \int} e^{- \beta E_{i}} \left[\sum_{\underset{s_j}{\underset{\mbox{\scriptsize  \textit{symm.}}}{\mbox{\scriptsize \textit{mixed}}}}} c_{s_j} \psi_{E_i}^{s_{j}}(1,\cdots,N)\right]\left[\displaystyle{\sum_{\underset{s_j}{\underset{\mbox{\scriptsize  \textit{symm.}}}{\mbox{\scriptsize \textit{mixed}}}}}} c_{s_j} \psi_{E_i}^{*s_{j}}(1,\cdots,N)\right] d1 \cdots dN.
\end{array}
\end{equation}

Relation \eqref{reves} is exact. Its relevance is of a theoretical character, as in general it will not be possible to calculate $Z_{M}(\beta, V,N)$. For the 2-particle system, \eqref{dosca} reduces to:

$$Z(\beta, V, 2)=\frac{Z_{S}+Z_{A}}{2},$$
as $Z_{M}(\beta, V,2)$ is zero. In Appendix A we calculate this expression for the 3-particle system.


\subsection{Symmetric functions versus symmetrized functions}

One of the crucial points in our view is noting the difference between the (originally) symmetric functions $\psi_{E_i}$ and the symmetrized functions $\psi_{E_i}^{S/A}$, constructed according to the rule \eqref{psisa}.
	
	In the case of 2 particles, if $\psi_{E_i}(1,2)$ is originally symmetric/antisymmetric then

$$\psi_{E_i}^{S/A}(1,2)=\sqrt{2}\psi_{E_i}(1,2) \qquad \mbox{and} \qquad \psi_{E_i}^{A/S}(1,2)=0.$$

Similarly, if an original $\psi_{E_i}(1, \cdots,N)$ is symmetric (or antisymmetric):

\begin{equation} \label{tomaya}
\psi_{E_i}^{S/A}(1,\cdots,N)=\sqrt{N!} \psi_{E_i}(1,\cdots,N).
\end{equation}

This expression gives the relation between the original symmetric wave functions $\psi_{E_i}$'s and the symmetrized ones $\psi_{E_i}^{S/A}$'s. Imagine now a Hamiltonian for which every original $\psi_{E_i}$ is symmetric/antisymmetric. Then, in \eqref{reves} only the first/second term in the right side would survive:

$$Z(\beta, V,N)=\frac{Z_{S/A}(\beta, V,N)}{N!}.$$
As:

\begin{equation} \label{tomayados}
Z_{S/A}(\beta, V,N)=  Tr \{\hat{\rho}_{S/A}(1, \cdots, N;1, \cdots, N) \}, 
\end{equation}
and according to \eqref{tomaya}

\begin{equation} \label{tomayatres}
Tr\{\hat{\rho}_{S/A} (1, \cdots, N;1, \cdots, N)  \}=N! Tr\{\hat{\rho}(1, \cdots, N;1, \cdots, N)  \},
\end{equation}
finally:
\begin{equation} \label{tomayatres}
Z(\beta, V,N)=  Tr \left\{\hat{\rho}(1, \cdots, N;1, \cdots, N) \right\}, 
\end{equation}
which is perfectly coherent with the definitions given above. Therefore, the factor $N!$ in \eqref{reves} is precisely what compensates the factor coming from the process of symmetrization in \eqref{tomaya}.

\section{Final Remarks}

The addition of the factorial of $N$ in the canonical partition function of an ideal gas has been the subject of hundreds of papers since Sackur and Tetrode introduced it in the early 1910's \cite{Monaldi}. Even after its quantum-mechanical justification in the 1930's (as far as we know, the first was due to G. E. Uhlenbeck and L. Gropper in 1932 \cite{Uhlenbeck}), many collateral issues have been (and still are being) discussed, mostly related to how we understand indistinguishability and identity. 

In this paper, we have focused on one specific point which is closely related to those fundamental issues, but which we have tackled in a purely mathematical way. Let us recapitulate the expressions that constitute the core of our paper:

\begin{equation} \label{pertu}
Z_{S/A}(\beta, V,N)= \displaystyle{\int \cdots \int}  \sum_{P} (\pm 1)^{P} \hat{\rho}(1, \cdots, N; P[1, \cdots, N]) d1 \cdots dN.
\end{equation}

\begin{equation} \label{finito}
\hat{\rho}=\frac{\hat{\rho}_{S}+ \hat{\rho}_{A}}{N!}
+ \hat{\rho}_{M}.
\end{equation}
These two relations are exact. Finally, in the classical limit we have obtained:
\begin{equation} \label{veri}
Z_{S/A}(\beta, V,N)\approx Z(\beta, V,N).
\end{equation}
This result differs from the generally admitted in a numerical factor $N!$ \cite{Feynman, Munster, Pathria, Huang}:

\begin{equation} \label{tpip}
^{\dagger}Z_{S/A}(\beta, V,N)\approx \frac{Z(\beta, V,N)}{N!} .
\end{equation}
It is said that the factor $N!$ is added in the classical treatment in an arbitrary way in order to obtain extensive expressions, for instance, for the ideal gas. It is also believed that this factor appears naturally in the quantum mechanical treatment. We have showed that this is no the case. 

We think that extensivity represents one of the most drastic illustrations of the fundamental inconmensurability between Thermodynamics and Mechanics. Nothing analogous to extensivity can be defined within the frame of Mechanics. This is correct not only for Classical but also for Quantum Mechanics, to which extensivity is as allien as to its predecessor. Therefore, we think that the absence of $N!$ in \eqref{veri} fits better with the role extensivity has in Thermodynamics. What we have shown is that a proper treatment cancels that factor in \eqref{tpip} coming from normalization (nothing more far related to extensivity) with another normalization factor, and prevents the quantum partition function from yielding extensive thermodynamic quantities. Hence, according to our result, the factorial of $N$ should be added in Classical Mechanics as well as in Quantum Mechanics. In fact, in a future paper we will argue that it would be more convenient to divide the partition function by $N^N$ instead of $N!$

\begin{appendix}

\section{3-particle system}

We are going to exemplify the concepts and results showed in sections 2 and 3 for the 3-particle system. In this case, the Hilbert space decomposes in four subspaces \cite{Galindo}:

$$\mathcal{H}^{(3)}=\mathcal{H}_{S}\oplus \mathcal{H}_{A} \oplus \mathcal{H}_{s_1} \oplus \mathcal{H}_{s_2}.$$
Therefore, we cannot write any eigenfunction in terms of symmetrized and antisymmetrized functions only. We have to consider eigenfunctions that have other symmetries. These are the functions of the new basis: 

\begin{equation} \label{tressimi}
\begin{array} {c}
\psi_{E_i}^{S}(1,2,3)=\frac{1}{\sqrt{6}} \left[\psi_{E_i}(1,2,3) +\psi_{E_i}(1,3,2)+\psi_{E_i}(2,3,1) + \right. \\ \left. +\psi_{E_i}(2,1,3)+\psi_{E_i}(3,1,2)+\psi_{E_i}(3,2,1) \right]
\end{array}
\end {equation}

\begin{equation} \label{tresanti}
\begin{array} {c}
\psi_{E_i}^{A}(1,2,3)= \frac{1}{\sqrt{6}} \left[\psi_{E_i}(1,2,3) -\psi_{E_i}(1,3,2)+\psi_{E_i}(2,3,1)- \right. \\ \left. -\psi_{E_i}(2,1,3)+\psi_{E_i}(3,1,2)-\psi_{E_i}(3,2,1) \right]
\end{array}
\end {equation}

\begin{equation} \label{M1}
\begin{array} {c}
\psi_{E_i}^{s_1}(1,2,3)= \frac{1}{2\sqrt{3}} \left[2 \psi_{E_i}(1,2,3) -\psi_{E_i}(1,3,2)+2\psi_{E_i}(2,1,3) - \right. \\ \left. - \psi_{E_i}(2,3,1) - \psi_{E_i}(3,1,2) - \psi_{E_i}(3,2,1) \right]
\end{array}
\end {equation}

\begin{equation} \label{M2}
\begin{array} {c}
\psi_{E_i}^{s_2}(1,2,3)=\frac{1}{2} \left[\psi_{E_i}(1,3,2) -\psi_{E_i}(2,3,1)+\psi_{E_i}(3,1,2) - \psi_{E_i}(3,2,1)\right]
\end{array}
\end {equation}

\begin{equation} \label{M3}
\begin{array} {c}
\psi'^{s_1}_{i}(1,2,3)= \frac{1}{2\sqrt{3}} \left[2 \psi_{E_i}(1,2,3) +\psi_{E_i}(1,3,2)-2\psi_{E_i}(2,1,3) -\right. \\ \left. - \psi_{E_i}(2,3,1) - \psi_{E_i}(3,1,2) + \psi_{E_i}(3,2,1) \right]
\end{array}
\end {equation}

\begin{equation} \label{M4}
\begin{array} {c}
\psi'^{s_2}_{i}(1,2,3)= \frac{1}{2} \left[\psi_{E_i}(1,3,2) +\psi_{E_i}(2,3,1)-\psi_{E_i}(3,1,2) - \psi_{E_i}(3,2,1)\right].
\end{array}
\end{equation}
Eigenfunctions \eqref{M1}, \eqref{M2}, \eqref{M3} and \eqref{M4} represent mixed symmetries, and they have to be grouped in pairs, \eqref{M1} and \eqref{M2}, and \eqref{M3} and \eqref{M4}: to generate its own subspace they require two independent functions instead of one. They are neither symmetric nor antisymmetric, but under consecutive permutations the generated subspace remains stable (after successive permutations, the obtained functions belong to the same subspace).

With a little calculation we can find that we can express $\psi_{E_i}(1,2,3)$ as follows:

$$\psi_{E_i}(1,2,3)=\frac{1}{\sqrt{6}}\left[ \psi^{S}_{E_{i}}(1,2,3) + \psi^{A}_{E_{i}}(1,2,3) \right]+ \frac{2 \sqrt{3}}{6} \left[ \psi^{s_1}_{E_{i}}(1,2,3)+  \psi'^{s_1}_{E_{i}}(1,2,3) \right].$$

Then, again, the join set formed by symmetrized and antisymmetrized functions do not constitute a complete set. We need more functions (besides \eqref{tressimi} and \eqref{tresanti}) to construct the new basis. 

The elements of the density matrix are:

\begin{equation} \label{mtzdscthree}
\begin{array} {c}
\hat{\rho}(1, 2, 3;1', 2', 3')

= \\ = \displaystyle{\frac{\left[ \hat{\rho}_S(1,2,3; 1', 2', 3') + \hat{\rho}_A (1, 2,3; 1',2', 3') \right]}{3!}} + \hat{\rho}_{M}(1,2,3;1',2',3'),
\end{array}
\end{equation}
where

\begin{equation}
\begin{array} {c}
\hat{\rho}_{M}(1,2,3;1',2',3')= \displaystyle{\frac{1}{6}} \sum_{\underset{\mbox{\scriptsize  \textit{states only}}}{\mbox{\scriptsize \textit{sym. \& antisym.}}}} e^{-\beta E_i} \left[ \psi_{E_i}^{S}(1,2,3) \psi^{*A}_{E_{i}}(1',2',3') + \psi^{A}_{E_{i}}(1,2,3)  \psi^{*S}_{E_{i}}(1',2',3') \right] +\\
+ \displaystyle{\frac{\sqrt{2}}{6}} \sum_{\underset{\mbox{\scriptsize  \textit{states}}}{\mbox{\scriptsize \textit{sym. \& antisym \& mix sym.}}}} \left\{ \left[ \psi_{E_i}^{S}(1,2,3)  + \psi_{E_i}^{A}(1,2,3) \right]  \left[ \psi^{*s_1}_{E_{i}}(1',2',3')+  \psi'^{*s_1}_{E_{i}}(1',2',3') \right] \right\} + \\
\displaystyle{\frac{1}{3}} \sum_{\underset{\mbox{\scriptsize  \textit{states only}}}{\mbox{\scriptsize \textit{mix. sym.}}}} \left\{ \left[ \psi^{s_1}_{i}(1,2,3)+  \psi'^{s_1}_{i}(1,2,3) \right]  \left[ \psi^{*s_1}_{E_{i}}(1',2',3')+  \psi'^{*s_1}_{E_{i}}(1',2',3') \right] \right\}.
\end{array}
\end{equation}
And the partition function:
\begin{equation}
\begin{array} {c}
Z(\beta, V, 3)= Tr\{\hat{\rho}(1,2,3;1,2,3)\} = \displaystyle {\frac{1}{3!}} \left[Z_{S}(\beta, V, 3) + Z_{A}(\beta, V,3)  \right]+  Z_{M}(\beta, V,3),
\end{array}
\end{equation}
where:

$$Z_{M}=\frac{1}{3} \left\{\int \int \int e^{- \beta E_{i}}d1 d2 d3 {|\psi_{E_i}^{s_{1}}(1,2,3)|^{2} + |\psi_{E_i}^{s'_{1}}(1,2,3)|^{2}}    \right\}.$$


\section{The Closure Property}

One of the first derivations of the classical limit of the canonical partition function was proposed by John G. Kirkwood in 1933, although he did not perform it explicitly \cite{Kirkwood}. Ironically, Kirkwood obtained the same result as ours \eqref{starring}, but only because he did not follow exactly his own proposal. Eventually, M\"unster as well as Grossman did perform it in detail \cite{Munster, Grossmann}. 

We are not going to reproduce here the step-by-step derivation, which is splendidly exposed in those books, but limit ourselves to point out the key point in the reasoning. It relies on an incorrect use of the closure condition. In the quoted derivations, we find, for the subspaces of symmetric (or antisymmetric) functions, the wrong relation:

\begin{equation}  \label{clausure}
^{\dagger}\sum_{\underset{\mbox{\scriptsize  \textit{states only}}}{\mbox{\scriptsize \textit{sym./antisym.}}}} \psi_{E_i}^{S/A}(1,\cdots,N) \psi_{E_i}^{*S/A}(1',\cdots,N')=\delta(1-1')\cdots \delta(N-N').
\end{equation}
This is the well-known closure condition, but applied in an improper way, because it should be applied only to complete sets of solutions $\psi_{E_i}$. As we have stated and showed above, neither symmetrized nor antisymmetrized functions (nor their joint set) constitute a complete set. 

The closure property is obtained \cite{Schiff} starting from the identity:
\begin{equation}  
 \hat{I}= \sum_{\underset{\mbox{\scriptsize  \textit{states}}}{\mbox{\scriptsize \textit{all}}}} \left. |\psi_{E_i} \right\rangle \left\langle \psi_{E_i}| \right. ,
 \end{equation}
and then calculating its matrix elements:

\begin{equation} 
\left\langle  1, \cdots, N \right. | \hat{I} | \left. 1', \cdots, N' \right\rangle ,
 \end{equation}
which are:

\begin{equation} \label{deltees}
\displaystyle{\sum_{\underset{\mbox{\scriptsize  \textit{states}}}{\mbox{\scriptsize \textit{all}}}}} \psi_{E_i} (1, \cdots, N) \psi^{*}_{E_i} (1', \cdots, N') = \delta(1-1') \cdots \delta(N-N').
\end{equation}

For the 2-particle system, where the subsystems formed by symmetrized and antisymmetrized functions generate the same Hilbert space as the original $\psi_{E_i}$'s, the particularization of \eqref{deltees} yields:

\begin{equation} \label{clausurerdospar}
\begin{array} {c}
\displaystyle{\sum_{\underset{\mbox{\scriptsize  \textit{states only}}}{\mbox{\scriptsize \textit{sym./antisym.}}}}} \psi_{E_i}^{S/A}(1,2) \psi_{E_i}^{*S/A}(1',2') = \\ =\displaystyle{\frac{1}{2}  \sum_{\underset{\mbox{\scriptsize  \textit{states}}}{\mbox{\scriptsize \textit{all}}}} [\psi_{E_i}(1,2) \pm \psi_{E_i}(2,1)] [\psi_{E_i}^{*}(1',2') \pm \psi_{E_i}^{*}(2',1')] }= \\ = \displaystyle{\frac{1}{2}  \displaystyle{\sum_{\underset{\mbox{\scriptsize  \textit{states}}}{\mbox{\scriptsize \textit{all}}}}}}  [\psi_{E_i}(1,2) \psi^{*}_{E_i} (1',2') + \psi_{E_i}(2,1) \psi^{*}_{E_i} (2',1') \pm \\ \pm \psi_{E_i}(1,2) \psi^{*}_{E_i} (2',1') \pm  \psi_{E_i}(2,1) \psi^{*}_{E_i} (1',2')].
\end{array}
\end{equation}
And according to \eqref{deltees} this is:

\begin{equation}
\frac{1}{2} \left[\delta(1-1')\delta(2-2')+\delta(2-2')\delta(1-1') \pm \delta(1-2')\delta(2-1') \pm \delta(2-1')\delta (1-2') \right] .
\end{equation}
Finally:
\begin{equation}
\displaystyle{ \displaystyle{\sum_{\underset{\mbox{\scriptsize  \textit{states only}}}{\mbox{\scriptsize \textit{sym./antisym.}}}}} \psi_{E_i}^{S/A}(1,2) \psi_{E_i}^{*S/A}(1',2') d1' d2'}=\delta(1-1')\delta(2-2') \pm  \delta(1-2')\delta(1'-2).
\end{equation}

For the case $N> 2$, recalling the construction rule for the symmetrized functions, \eqref{psisa},we obtain:

\begin{equation} \label{tancant}
\begin{array} {c}
\displaystyle{ \sum_{\underset{\mbox{\scriptsize  \textit{states only}}}{\mbox{\scriptsize \textit{sym./antisym.}}}}} \psi_{E_i}^{S/A}(1, \cdots, N) \psi_{E_i}^{*S/A} (1', \cdots, N')= \\
= \displaystyle{ \frac{1}{N!}} \displaystyle{\sum_{\underset{\mbox{\scriptsize  \textit{states}}}{\mbox{\scriptsize \textit{all}}}}} \left\{ \displaystyle{\sum_{P}} (\pm 1)^{P} \psi_{E_i} (P[1, \cdots, N]) \right\} \left\{\displaystyle{\sum_{P'}} (\pm 1)^{P'} \psi^{*}_{E_i} (P'[1', \cdots, N']) \right\}.
\end{array}
\end{equation}
Similarly to the 2-particle example, the first factor as well as the second consists of $N!$ addends, $N!$ permutations. Each permutation $P$ has its equivalent $P'$, and there are $N!$ per each. Then, appealing again to the closure condition \eqref{deltees} for the original functions, we obtain (here $\textbf{Q}$ stands for $(1, \cdots,N$)):

\begin{equation} \label{clausurerdos}
\sum_{\underset{\mbox{\scriptsize  \textit{states only}}}{\mbox{\scriptsize \textit{sym./antisym.}}}} \psi_{E_i}^{S/A}(1,\cdots,N) \psi_{E_i}^{*S/A}(1',\cdots, N') d1 \cdots dN=\sum_{P} (\pm 1)^{(P)} \delta(\textbf{Q}-P[\textbf{Q'}]),
\end{equation}
which is different from \eqref{clausure}. 

\section {Overcounting Symmetry}

The method we have developed has a general character, that is, its applicability is not restricted to a simple kind of system. However, in many textbooks what we find is the particularization of result \eqref{permi2} to ideal gases. Hence, we would like to make some comments on that wide-spread derivation.

	In order to do that, let us go back to probably the first derivation of this kind by G. Uhlenbeck and L. Gropper in 1932 \cite{Uhlenbeck}. Our remarks are also valid for many different derivations still usual nowadays \cite{Pathria, Huang}.
	
	Uhlenbeck and Gropper operate with the eigenfunctions for two particles in a box (of one dimension):

\begin{equation} \label{gon}
\psi_{n_{1}, n_{2}}(q_1, q_2)=\frac{2}{L} \sin \frac{n_1 \pi q_1}{L} \sin \frac {n_2 \pi q_2}{L}
\end{equation}
($q_1$ and $q_2$ are the coordinates of the first and the second particle). The matrix elements are:

\begin{equation} \label{doblebe}
\sum_{n_{1}, n_{2}} e^{-\beta E_{n_{1}, n_{2}}} \psi_{n_{1}, n_{2}} (q_1, q_2) \psi_{n_{1}, n_{2}} (q'_1, q'_2) ,
\end{equation}
with:

$$E_{n_{1}, n_{2}}=\frac{\alpha \pi^{2}}{L^{2}} (n^{2}_{1} + n^{2}_{2}).$$
%
The trace can be easily calculated as:

\begin{equation} \label{uhlbz}
\frac{4}{L^2}\sum_{n_1=0}^{\infty} \sum_{n_2=0}^{\infty} e^{-\frac{\alpha \pi^{2}}{L^{2}} (n_1^2+n_2^2)} \sin^2 \frac{n_1 \pi q_1}{L} \sin^2 \frac{n_2 \pi q_2}{L},
\end{equation}
where the sums over $n_1$ and $n_2$ are completely independent of each other.

Next, using symmetrized functions:

\begin{equation} \label{symys}
\psi^{S/A}_{n_{1}, n_{2}}(q_1, q_2)= \frac{1}{\sqrt{2}} \frac{2}{L} \left\{\sin \frac{n_1 \pi q_1}{L}\sin \frac{n_2 \pi q_2}{L} \pm \sin \frac{n_1 \pi q_{2}}{L}\sin \frac{n_2 \pi q_{1}}{L}\right\},
\end{equation}
they define the corresponding sum for the case of Bose-Einstein ($+$) and Fermi-Dirac ($-$):

\begin{equation} \label{uhlbe}
\frac{2}{L^2}\sum \sum_{n_1\geq n_2} e^{-\frac{\alpha \pi^{2}}{L^{2}} (n_1^2+n_2^2)}\left(\frac{1}{2} \right)^{\delta_{n_1,n_2}} \left\{\sin \frac{n_1 \pi q_{1}}{L} \sin \frac{n_2 \pi q_{2}}{L} \pm \sin \frac{n_1 \pi q_{2}}{L} \sin \frac{n_2 \pi q_{1}}{L} \right\}^{2}
\end{equation}
($\delta_{n_1,n_2}$ is Kronecker delta). Let us compare equations \eqref{uhlbz} and \eqref{uhlbe} carefully. In \eqref{uhlbz} there are no symmetrized functions. On the contrary, in \eqref{uhlbe} Uhlenbeck and Gropper consider only symmetrized functions, built starting from the unsymmetrized ones. Our point is that as they are still in the original basis of unsymmetrized functions, they should remain there to work out the sum. In other words, in terms of our illustration with the 2-particle system in section 3.1, we do know how to sum \eqref{correctavvv}, not \eqref{correctavv}. 

In contrast, Uhlenbeck and Gropper add condition $n_1\geq n_2$. This condition would mean that we are able to distinguish particles (the first one being the particle with the greatest energy, for instance), which is exactly what we want to avoid. Once we have symmetrized the eigenfunctions, the distinction between states of the type $(n_1,n_2)$ and $(n_2,n_1)$ is redundant. In fact, it is the use of \eqref{symys} which allows us to perform the sum in \eqref{uhlbe} without restrictions. 

It is because of that that we think that Gropper and Uhlenbeck \textit{overcount} the symmetry: they add the condition $n_1\geq n_2$ once the linear combination in \eqref{symys} had already made that distinction unnecessary (meaningless). 

	According to our view, the corrected sum should be, instead of \eqref{uhlbe}:

\begin{equation} \label{uhlbec}
\frac{2}{L^2}\sum_{n_1=0}^{\infty} \sum_{n_2=0}^{\infty} e^{-\frac{\alpha \pi^{2}}{L^{2}} (n_1^2+n_2^2)} \left\{\sin \frac{n_1 \pi q_{1}}{L} \sin \frac{n_2 \pi q_{2}}{L} \pm \sin \frac{n_1 \pi q_{2}}{L} \sin \frac{n_2 \pi q_{1}}{L} \right\}^{2}.
\end{equation}

We obtain the same result if we apply our formula \eqref{permi2}:

$$\hat{\rho}_{S/A}(1,2;1,2)=\hat{\rho}(1,2;1,2) \pm \hat{\rho}(1,2;2,1),$$
and recalling that $\hat{\rho}(1,2;1,2)$ is \eqref{uhlbz}.

To obtain the partition function we have to perform the sum for $n_1$ and $n_2$ and later on integrate expressions \eqref{uhlbz} and \eqref{uhlbec} along $q_1$ and $q_2$. Uhlenbeck and Gropper show masterly how to perform this calculation. Again, our final result differs from theirs in the factorial of $N$.

\end{appendix}

\ack

Partial financial support from the Ministerio de Ciencia e Innovaci\'on under Contract No. FIS2009-09689 is acknowledged. We thank Anthony Duncan and Gonzalo Bermejo for his assistance during the preparation of the paper.

\end{document}